\documentclass[aps,preprint]{revtex4}%
\usepackage{amsfonts}
\usepackage{amsmath}
\usepackage{amssymb}
\usepackage{graphicx}%
\providecommand{\U}[1]{\protect\rule{.1in}{.1in}}

\begin{document}
\preprint{HEP/123-qed}
\title{Electron channeling, de Broglie's clock and the relativistic dynamical time operator}
\author{M. Bauer}
\affiliation{Instituto de F\'{\i}sica, Universidad Nacional Aut\'{o}noma de M\'{e}xico}
\affiliation{A.P. 20-364, 01000 M\'{e}xico, D.F., MEXICO }
\pacs{PACS number}

\begin{abstract}
Electron channeling in silicon crystals has brought forward the possibility of
having detected a particle's "internal clock", as an intrinsic oscillation
with de Broglie's frequency. The transmission probability along a major axial
direction is reduced with respect to neighboring angles, except for a sharp
peak at the atomic row direction. The pattern observed is a "W" instead of a
"U". This central peak is attributed to a process known as "rosette motion",
in which the crossing of successive atoms would be related to the de Broglie
frequency. A classical multiple scattering calculation found that, to
represent the experimental results, the interaction frequency had to be about
twice the de Broglie's clock frequency; that is, the "Zitterbewegung"
frequency. In the present paper, the observed characteristics of this process
are shown to be consistent with a free particle quantum mechanical motion
described by Dirac's Hamiltonian, albeit with an effective mass resulting from
the interaction with the crystal atoms. The introduction of a self-adjoint
dynamic time operator provides the connection with an internal "system time",
the de Broglie clock.

\end{abstract}
\date[Date text]{date}
\maketitle

\section{Introduction}

Electron channeling in silicon crystals\cite{Catillon,Gouanere} has brought
forward the possibility of having detected a particle's \textquotedblleft
internal clock\textquotedblright, as an intrinsic oscillation whose frequency
is given by de Broglie's daring association $h\nu=m_{0}c^{2}$, where $h$ is
Planck's constant, $m_{0}$ is the particle rest mass and $c$ the speed of
light in vacuum.\cite{Broglie} More recently, a clock linked to this relation
has been demonstrated using an optical frequency to self-reference a
Ramsey-Bord\'{e} atom interferometer.\cite{Lan}

The channeling experiments in which the electrons are aligned along a major
axial direction of a thin single crystal, do exhibit a reduced transmission
probability with respect to neighboring angles, except for a sharp peak at the
atomic row direction. The pattern observed is a \textquotedblleft
W\textquotedblright\ instead of a \textquotedblleft U\textquotedblright. This
central peak is attributed to a process known as \textquotedblleft rosette
motion\textquotedblright, which results in a reduction of the multiple
scattering effects for electrons moving parallel to a string of atoms, with a
momentum such that they pass atoms with a frequency equal to the de Broglie
frequency. The expected consequence is a higher transmittivity relative to
closely nearby directions and momenta. A phenomenological calculation by a
Montecarlo method was carried out, in which classical mechanics was used to
describe the electron motion. It was found however that to represent the
experimental results, the interaction frequency had to be about twice the de
Broglie's clock frequency; that is, closer to the \textquotedblleft
Zitterbewegung\textquotedblright\ frequency, which appears in Dirac's
relativistic formulation of quantum mechanics.

In the present paper, the observed characteristics of this process are shown
to be consistent with a free particle motion described by a Dirac Hamiltonian,
albeit with an effective mass resulting from the interaction with the crystal
atoms. The introduction of a dynamic time operator\cite{Bauer} provides the
connection with an internal system time, the de Broglie clock.

\section{The free particle Dirac Hamiltonian as a symmetry operation}

Consider the free particle Dirac Hamiltonian
\[
H=c\mathbf{\boldsymbol{\alpha}.p}+\beta m_{0}c^{2}%
\]
where \textbf{$\alpha$}$=(\mathbf{\alpha}_{x},\mathbf{\alpha}_{y}%
,\mathbf{\alpha}_{z})$ and $\beta$ are the Dirac matrices. Recalling that the
infinitesimal $(\epsilon\ll1)$ unitary operator $S(\epsilon)=e^{\{i\epsilon
p/\hbar\}}$ acting on a position eigenstate yields a displaced eigenstate,
namely\cite{Messiah}:%
\[
S(\epsilon)|x>=e^{\{i\epsilon p/\hbar\}}|x>==[1+(i\epsilon p/\hbar)+%
\frac12
(i\epsilon p/\hbar)%
{{}^2}%
+.....]|x>=|x+\epsilon>,
\]
it follows that the infinitesimal unitary operator $(\tau\ll1)$:%
\begin{align}
U(\tau) &  =e^{\{-i\tau H/\hbar\}}=e^{\{-i\tau\{c\mathbf{\boldsymbol{\alpha
}.p}+\mathbf{\beta}m_{0}c^{2}\}/\hbar\}}=\nonumber\\
&  =[1+(i\tau\mathbf{\boldsymbol{\alpha}.p}/\hbar c)+%
\frac12
(i\tau\mathbf{\boldsymbol{\alpha}.p}/\hbar c)%
{{}^2}%
+.....]exp^{(-i\tau m_{0}c^{2}/\hbar)}%
\end{align}
induces, in configuration space, a position displacement by an amount
$\delta\mathbf{r}=\tau c$\textbf{$\boldsymbol{\alpha}$} and a phase shift
$\delta\phi=\beta(\tau m_{0}c^{2}/\hbar)$, i.e.:%
\begin{equation}
\Phi(\mathbf{r})=<\mathbf{r}\mid\Phi>=e^{i\phi}\varphi(\mathbf{r})\rightarrow
e^{i(\phi+\mathbf{\delta}\phi)}\varphi(\mathbf{r}+\tau
c\mathbf{\boldsymbol{\alpha}})(2).
\end{equation}
\qquad\ As $[S(\tau),H]=0$, the displaced wave function satisfies the same
Schr\"{o}dinger equation. $S(\tau)$ is thus a symmetry operation.

Averaging over a general positive energy wave packet (i.e., with $<\beta>=1$
but including both positive and negative energy eigenstates)
yields\cite{Greiner}:%
\begin{equation}
<\delta\mathbf{r}>=\tau<c\mathbf{\boldsymbol{\alpha}}>=\tau\{<c%
{{}^2}%
\mathbf{p}/E>+\ oscillating\ terms\ (Zitterbewegung)\},
\end{equation}
where $<c%
{{}^2}%
p/E>$ is the group velocity $v_{gp}=dE/dp$. Finite displacements are achieved
by repeated applications. The phase shift is seen to be related to the reduced
de Broglie frequency $(m_{0}c^{2}/\hbar)$.

\section{The dynamical time operator and electron channeling}

The introduction of the self-adjoint dynamical time operator:%
\begin{equation}
T=\mathbf{\boldsymbol{\alpha}.r}/c+\beta\tau_{0}%
\end{equation}
as a system observable\cite{Bauer}, establishes that the phase velocity
represents the change rate of the displacement with respect to the internal
system time:%
\begin{equation}
d\mathbf{r}(t)/dT(t)=\mathbf{v}_{ph},
\end{equation}
whereas the group velocity yields the change rate with respect to the
laboratory time $t$\cite{Messiah,Greiner}, i.e.:%

\begin{equation}
d\mathbf{r}(t)/dt=\mathbf{v}_{gp}.
\end{equation}
In Eq.(5) and in Eq.(6), the oscillating terms (Zitterbewegung) have been
omitted (they are not present in th case of wave packets of purely positive or
purely negative energy eigenfunctions\cite{Messiah,Greiner}). Phase and group
velocities satisfy the relation $\mathbf{v}_{ph}\mathbf{v}_{gp}=c^{2}$.

Consider now a crystal with $d$ as the separation between atoms. The system
time lapse needed for a fixed phase point to achieve such displacement will
be, from Eq.(5):%
\begin{equation}
\Delta T=d/v_{ph}%
\end{equation}
and the corresponding phase shift at $d$ will be increased by:%
\begin{equation}
\Delta\varphi=(\Delta T)m_{0}c^{2}/\hbar=(d/vph)m_{0}c^{2}/\hbar\mathbf{.}%
\end{equation}
An increment equal to $n\pi\ (n=1,2,\ldots)$ results in the same amplitude
absolute value (odd $n$ only changes the sign). The corresponding system time
lapses are:%
\begin{equation}
(\Delta T)_{n\pi}=(n\pi)(\hbar/m_{0}c^{2})=(n/2)(h/m_{0}c^{2}),
\end{equation}
that is, $(n/2)$ times the de Broglie period. The corresponding phase velocity
must then be:%
\begin{equation}
(v_{ph})_{n\pi}=(2/n)d(m_{0}c^{2}/h).
\end{equation}
As $v_{ph}=E/p$, it follows that%
\begin{align}
E_{n\pi}/c  &  =p[(v_{ph})_{n\pi}/c]=(m_{0}\gamma c)[(v_{ph})_{n\pi
}/c]\nonumber\\
&  =(m_{0}c^{2}\gamma)[(v_{ph})_{n\pi}/c]/c,
\end{align}
where $\gamma$ is the relativistic Lorentz factor $[1-(v_{gp}/c)^{2}%
]^{-1/2}=[1-(c/v_{ph})^{2}]^{-1/2}.$

\section{The electron channeling experiment}

In the electron channeling experiment performed\cite{Catillon},
$d=3.84\ \mathring{A}=3.84\ 10^{5}\ F$ and therefore $[(v_{ph})_{n\pi
}/c]=(2/n)158.265$ and $[(v_{gp})_{n\pi}/c]=(n/2)(158.265)^{-1}$. With this
value of the group velocity, $\gamma$ is practically equal to $1$ so no
relativistic correction to the electron mass needs to be considered. Then:%
\begin{equation}
E_{n\pi}/c=(2/n)(0.511)(158.265)/c=(2/n)\ 80.873\ MeV/c.
\end{equation}

When $n=2,$ $E_{2\pi}/c=80.873\ MeV/c,$ close to the observed rosette motion
resonance at $81.1\ MeV/c$. This corresponds to a phase shift of $2\pi$ and,
according to Eq.(10), to a phase velocity:%
\begin{equation}
v_{ph}=d(m_{0}c^{2}/h).
\end{equation}
i.e., the distance $d$ between atoms multiplied by the de Broglie frequency.

To be noted from Eq.(13) is that the same phase velocity would result from an
inter atomic distance $d/2$ times the Zitterbewegung frequency $2m_{0}c^{2}/h$
without modifying the resonance energy $E_{2\pi}$ from Eq.(12). These are the
conditions under which the classical multiple scattering calculation of Ref.1
agrees with the experiment. This is also in agreement with the improvement
found in Lindhard's continuum model when one considers the transverse energy
to be conserved only when measured on the transverse planes located midway
between atoms in a string.\cite{Lindhard,Gemmell}

Also to be noted on the other hand is the following. The value $n=1$
corresponds to a phase shift of $\pi$. This changes the sign of the wave
function but the absolute value remains the same at each crossing of a crystal
atom. In this case the phase velocity, (Eq.(10)), would be given by the
interatomic distance $d$ times the Zitterbewegung frequency $2m_{0}c^{2}/h$.
However, the rosetta motion resonance is then expected to be observed at twice
the energy, i.e., $161.746\ MeV$ (Eq.(12)), as noted in the earlier report of
the experiment in Ref.2.

Finally, the difference between theoretical and experimental momenta could be
accounted by considering that the electron's motion in the crystal is well
represented by the free particle Dirac Hamiltonian, albeit with an effective
mass $m^{\ast}$ instead of $m_{0}$ that takes into account the average
interaction of the electron with the crystal atoms\cite{Kittel,Green}.
Following this assumption through the above derivation yields ($m^{\ast}%
$/$m_{0}$)$^{2}$ $=(81.1)/(80.873)=1.0028$ and $m^{\ast}$ $=1.0014\ m_{0}$.

\section{Conclusion}

The resonant conditions observed in electron channeling experiments are shown
to be consistent with the quantum mechanical description provided by Dirac's
free motion Hamiltonian (albeit with an effective mass), and the associated
internal dynamical time operator. These conditions result from requiring: a)
the same absolute value of the wave function at each atomic crossing; b) the
association of the required phase velocity with the internal system time, as
obtained from the system self-adjoint time operator introduced as an
observable in a previous paper.\cite{Bauer}

\end{document}